\documentclass[amsmath,amssymb,aps,prb,reprint]{revtex4-1}

\usepackage{chemformula} 
\usepackage[T1]{fontenc} 

\usepackage{graphicx}
\usepackage{dcolumn}
\usepackage{lipsum}
\usepackage{tabu} 
\usepackage{mathtools}

\usepackage{braket}
\usepackage{color}
\usepackage[11pt]{moresize}
\usepackage{anyfontsize}
\usepackage{bbding}
\usepackage{physics}

\begin{document} 

\author{Khaled Mnaymneh}\email{khaled.mnaymneh@nrc-cnrc.gc.ca}
\author{Dan Dalacu}
\author{Joseph McKee}
\author{Jean Lapointe}
\author{Sofiane Haffouz}
\author{Philip J. Poole}
\author{Geof C. Aers}
\author{Robin L. Williams}

\affiliation{National Research Council Canada, Ottawa, Ontario, Canada, K1A 0R6.}

\title{Monolithic integration of single photon sources via evanescent coupling of tapered InP nanowires to SiN waveguides}



\begin{abstract}
We demonstrate a method to monolithically integrate nanowire-based quantum dot single photon sources on-chip using evanescent coupling. By deterministically placing an appropriately tapered III-V nanowire waveguide, containing a single quantum dot, on top of a silicon-based ridge waveguide, the quantum dot emission can be transferred to the ridge waveguide with calculated efficiencies close to 100\%. As the evanescent coupling is bidirectional, the source can be optically pumped in both free-space and through the ridge waveguide. The latter configuration provides a self-contained, all-fiber, single photon source suitable as a plug-and-play solution for applications requiring bright, on-demand single photons. Using InAsP quantum dots embedded in InP nanowire waveguides, we demonstrate coupling efficiencies to a SiN ridge waveguide of 74\% with a single photon purity of 97\%.
   
\end{abstract}


\maketitle 
\newpage

The integrated photonics platform has been identified\cite{Rudolph_APLP2017} as the most practical way to realize quantum technologies in the near-term\cite{Fiore_JO2017}. Critical to its widespread implementation is the design and integration of scalable quantum light sources\cite{Senellart_NN2017}. Integrated sources are attractive for experiments requiring complex photonic circuits such as linear optics quantum computing\cite{Thompson_2011} and quantum simulation\cite{Sparrow_NAT2018}. Integration is also relevant in the development of `plug and play' sources\cite{Xu_APL2008} for secure quantum communication, where standard waveguide-fiber coupling techniques\cite{Shi_PTL2014} can be utilized to reduce the complexity inherent to active alignment procedures required with free-space approaches.

Efforts to develop monolithically integrated sources include quantum dots coupled to ridge \cite{Schwartz_OE2016,Jons_JPD2015,Reithmaier_NL2015,Schnauber_NL2018}, nanobeam \cite{Prtijaga_APL2014,Kirsanske_PRB2017} and photonic crystal waveguides\cite{Schwagmann_APL2011,Laucht_PRX2012,Arcari_PRL2014,Sollner_NN2015} using the same material system as the quantum dot. Hybrid approaches where various two-level systems are coupled to silicon-based photonic circuits have also been investigated\cite{Hausmann_NL2012,Khasminskaya_NP2016,Ellis_APL2018}, inspired by techniques developed for laser integration\cite{Roelkens_LPR2010}. Zadeh \textit{et} \textit{al}, for example, have demonstrated a pick-and-place technique\cite{Zadeh_NL2016} where individual InAsP/InP quantum dot nanowires are transferred from the growth substrate to a silicon substrate. Subsequent processing is used to define SiN waveguides with the single emitters embedded within. Hybird integration based on evanescent coupling has also been demonstrated; in this case, devices having appropriate geometries are either fabricated on the III-V growth substrate and transferred to a silicon chip\cite{Kim_NL2017} or the entire epitaxial layer is transferred and processing is done on silicon\cite{Davanco_NC2017}. Both of the latter approaches utilized randomly nucleated dot ensembles and hence lacked the deterministic incorporation of single emitters used in Ref.~\citenum{Zadeh_NL2016}.

In this work we propose a hybrid integration method based on the evanescent coupling of a nanowire waveguide mode excited by a single quantum dot to an underlying ridge waveguide. Nanowire waveguides with well-controlled tapers\cite{nw_review} provide an ideal geometry for evanescent mode transfer. In contrast to previous work on nanowire on-chip integration\cite{Zadeh_NL2016,Elshaari_NC2017}, where the nanowire is embedded within the waveguide, the present approach can be used with pre-fabricated photonic integrated circuits via the nanomanipulator transfer technique described in Ref.~\citenum{Kim_NL2017}. In the work presented here however, single emitters are transferred deterministically. 

\begin{figure}
\begin{center}
\includegraphics*[width=8.5cm,clip=true]{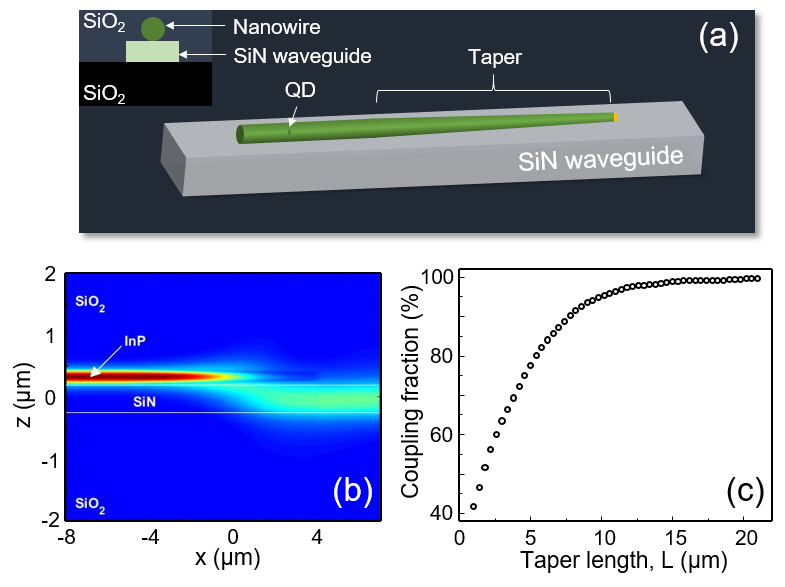}
\end{center}
\caption{(a) Schematic of the device for evanescently coupling the emission from an InP nanowire to a SiN waveguide. (b) Electric field $E_y$ of the fundamental waveguide mode in the coupled device. (c) Calculated coupling efficiency as a function of the nanowire taper length, $L$, for a ridge waveguide with $w=400$\,nm and $t=500$\,nm.}\label{fig1}
\end{figure}

A schematic of the device is shown in Figure~\ref{fig1}(a). It consists of a tapered InP nanowire placed on top of a SiN ridge waveguide. A single InAsP quantum dot is located in the untapered base of the nanowire which has a diameter $D_\mathrm{b}$ chosen to confine the fundamental $HE_\mathrm{11}$ mode\cite{nw_review}.  The upper section of the nanowire waveguide is tapered to evanescently transfer the $HE_\mathrm{11}$ mode to the SiN waveguide. We assume that the quantum dot is polarized to optimally couple to the TE mode of the nanowire-waveguide system.

In Figures~\ref{fig1}(b) and (c) we show the calculated coupling efficiency and its dependence on the taper length, $L$. The calculations were performed using eigenmode expansion methods\cite{lumerical} for a nanowire with base diameter $D_\mathrm{b}=250$\,nm tapered to a tip diameter $D_\mathrm{t}=100$\,nm over a length $L$. The nanowire is located on a SiN waveguide with a width $w$ and thickness $t$. The nanowire-waveguide device is encapsulated above and below with $5\,\mu$m of SiO$_2$ for waveguiding in the SiN/SiO$_2$ system. Devices with over 90\% transfer of the $HE_\mathrm{11}$ from the InP nanowire to the SiN waveguide after $10\,\mu$m of taper, approaching 100\% for longer tapers, are predicted. Calculations are shown for $w=400$\,nm and $t=500$\,nm although significant transfer efficiencies are predicted over a wide range of waveguide widths and thicknesses. We also note that this approach works equally well with the nanowires positioned beside the ridge waveguide.

\begin{figure}[h]
\includegraphics*[width=8.5cm,clip=true]{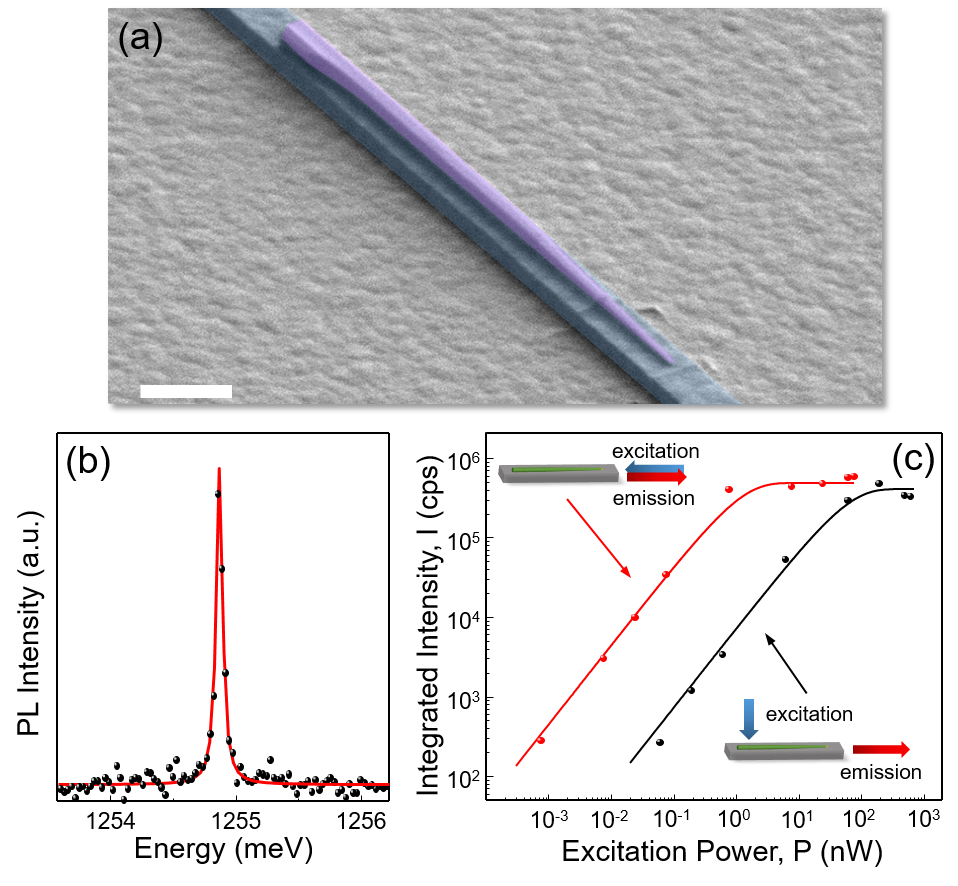}
\caption{(a) SEM image of a $9\,\mu$m long nanowire with a taper length $L=8\,\mu$m on a SiN ridge waveguide. Scale bar is $1\,\mu$m. (b) PL spectrum for excitation normal to the nanowire and collection from the chip edge. (c) Pulsed excitation power dependence of the integrated PL for free space (black symbols) and waveguide (red symbols) pumping, see insets. Solid lines are fits to a saturation function of the form $I\propto 1-e^{-P/P_{sat}}$.}
\label{fig2}
\end{figure}

To demonstrate the feasibility of an evanescently coupled on-chip nanowire source, we used the pick-and-place technique previously demonstrated in Ref.~\citenum{Zadeh_NL2016}. We use InAsP/InP nanowire quantum dot sources grown using a combined selective-area and vapour-liquid-solid (VLS) epitaxy approach\cite{Dalacu_NT2009} which have demonstrated single photon purities greater than 99\%\cite{Dalacu_NL2012} and near-transform-limited linewidths of less than $4\,\mu$eV\cite{Reimer_PRB2016}. Pre-selected nanowires were transferred from the InP growth substrate to a silicon substrate pre-coated with a $5\,\mu$m layer of SiO$_2$ and a $200$\,nm layer of SiN and patterned with gold alignment marks. The SiN waveguides were defined underneath the nanowire using electron-beam (e-beam) lithography and dry-etching. The device was then coated with a top $5\,\mu$m layer of SiO$_2$. Details of the device fabrication are given in Methods.

A scanning electron microscopy (SEM) image of a fabricated device prior to the top SiO$_2$ coating is shown in Figure~\ref{fig2}(a). The nanowire is well-aligned to the SiN waveguide and is tapered over $8\,\mu$m after an initial $1\,\mu$m untapered section which contains the quantum dot. To optically characterize the source, we excite above-band with either a continuous wave (cw) or a pulsed laser (see Methods) with the excitation oriented normal to the nanowire. The emission is collected off-chip using a cleaved fiber, butt-coupled to the SiN waveguide. Coupling losses are minimized by tapering the width of the SiN waveguide down to 200\,nm over a distance of $300\,\mu$m near the chip edge, adiabatically expanding the mode for matching to the single-mode fiber core\cite{Lipson_OL2003}. 

A photoluminescence (PL) spectrum from the source using cw excitation is shown in Figure~\ref{fig2}(b) and consists of a single peak identified at the neutral exciton from the power dependence of the intensity. A lorentzian fit to the peak gives a resolution-limited linewidth of $60\,\mu$eV and higher resolutions are required to verify that the integration method does not degrade the optical quality of the emitter. To estimate the evanescent coupling efficiency we use the integrated intensity, $I$, at a pulsed excitation power, $P_{sat}$, that saturates the transition. From Figure~\ref{fig2}(c), a count rate of $I=417$\,kHz was measured at $P_{sat}$ using a pulsed excitation rate of 80\,MHz. Taking into account an experimental system throughput of 1.85\% (see Table~\ref{table:1}), we estimate a count rate in the waveguide of 22.4\,MHz, giving a source efficiency of 28\%. Since the quantum dot emission is equally likely to emit in the forward and backward directions along the nanowire, one can estimate the evanescent coupling efficiency from twice the source efficiency (i.e. 56\%). If we account for 5\% emission into leaky modes\cite{nw_review} and 20\% emission into phonon sidebands at 4\,K\cite{Lodahl_RMP2015} (not included in the measured counts obtained using a lorentzian fit) the evanescent coupling efficiency is 74\%.

\begin{table}[h!]
\begin{center}
\caption{Experimental transmission and detection efficiency.}
\begin{tabular}{|l | c|} 
 \hline
\hspace{48mm} &\hspace{5mm}Throughput [\%]\hspace{5mm} \\  
 \hline
 Waveguide propagation & 69.2 \\
 Waveguide-fiber coupling& 45.7 \\
 Spectrometer & 11.7\\
 CCD detector efficiency & 50 \\
 \textbf{Total} & \textbf{1.86} \\
 \hline
\end{tabular}

The waveguide loss calculations are given in Supplementary Information. Spectrometer losses include fiber transmission and the monochromater (i.e. one grating, three mirrors). 
\label{table:1}
\end{center}
\end{table}

\begin{figure}
\begin{center}
\includegraphics*[width=8.5cm,clip=true]{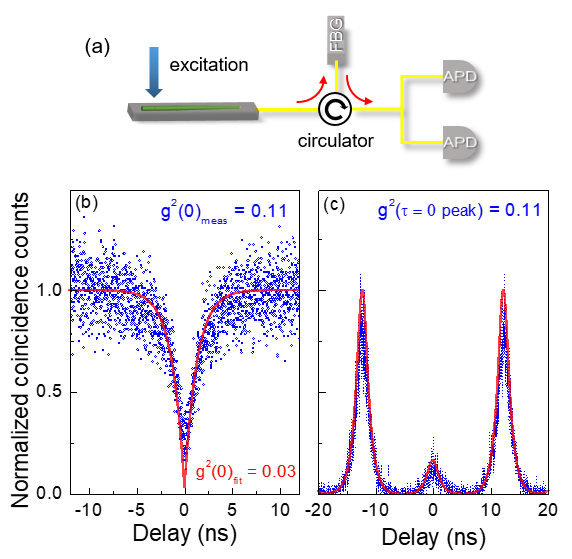}
\end{center}
\caption{(a) Schematic of the HBT set-up used to measure the photon emission statistics. Second-order correlation function measured using (b) cw and (c) pulsed excitation. Red curve in (b) is a fit after deconvolving a detector response of 675\,ps. Red curve in (c) is a fit using a re-excitation model.}
\label{fig3}
\end{figure}

To measure the photon statistics, an all-fiber Hanbury-Brown Twiss (HBT) interferometer was used. The emission from the source is directed to a pair of fiber-coupled Si avalanche photodiodes (APDs) after filtering through a tunable fiber Bragg grating (FBG), see Figure~\ref{fig3}(a) and Methods. The second-order correlation function, $g^2(\tau)$, under cw excitation measured at $P_{sat}$ is shown Figure~\ref{fig3}(b). The measured single-photon purity at zero delay is $1 -g^2(0) = 89\%$. The solid red line is corrected for the timing response of the APDs by deconvolving a fit to the data using a two-level second-order correlation function $g^2(\tau)=1-\mathrm{exp}(-|t|/\tau)$ and the detector time response of 675\,ps, where $\tau=1.3$\,ns is the radiative life-time of the transition. From the fit, a corrected single photon purity of 97\% is obtained. We conclude that the non-resonant pumping scheme and propagation through the waveguide has no major effect on the photon statistics.

Figure~\ref{fig3}(c) shows the $g^2(\tau)$ measured using pulsed excitation at $P_{sat}=800$\,nW. The measured single-photon purity in this case is also 89\%, determined from the ratio of the $\tau=0$ peak to the adjacent peaks. In contrast to the cw measurement, where single photon purity is determined from post-selected photons arriving at $\tau=0$, the pulsed measurement also provides information on photon coincidences at short delay times. The counts observed around zero delay suggests re-excitation of the nanowire quantum dot from carriers produced in the same excitation pulse, frequently observed using above-band excitation when pumping close to saturation\cite{Aichele_NJP2004,Santori_NJP2004}. We model the exciton re-excitation process from the band-edge as a competition between carrier recombination and dot capture after each exciton emission process. We find that a time of ~100 ps for both processes, shorter than both the exciton radiative lifetime and the pulse period, is required to reproduce the measured coincidences (solid red curve in Figure~\ref{fig3}(c)). 

\begin{figure}
\begin{center}
\includegraphics*[width=8.cm,clip=true]{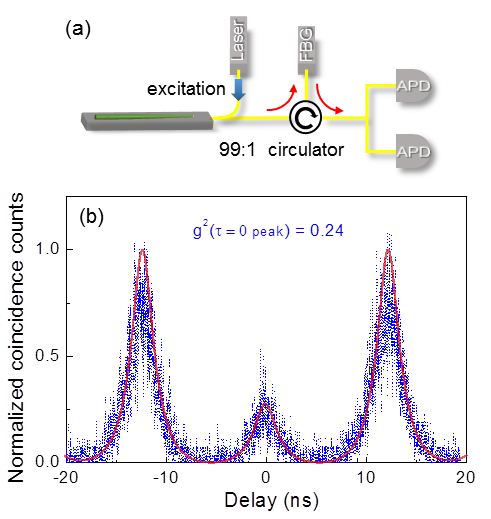}
\end{center}
\caption{(a) HBT measurement with the optical excitation fiber-coupled into the 1\% port of a 99:1 splitter where the splitter's transmission port is facet-coupled to the SiN waveguide. (b) Measured $g^2(\tau)$ using pulsed evanescent excitation. Red curve is a model fit to the data.}
\label{fig4}
\end{figure}

Since the evanescent coupling is bidirectional, the nanowire quantum dot can also be excited by pumping into the SiN waveguide. Figure~\ref{fig2}(c) shows the pump power dependence of the integrated PL from the quantum dot obtained using evanescent excitation (red symbols). In this case, optical excitation is provided by a fiber-coupled pulsed laser into the 1$\%$ port of a 99:1 fiber splitter that was facet-coupled to the SiN waveguide; the pump propagates within the SiN waveguide and evanescently excites the quantum dot. The reduced excitation power required to achieve saturation compared to free-space pumping is attributed to an increased absorption cross-section in the case of evanescent pumping. In Figure~\ref{fig4} we show a second-order correlation measurement from the evanescently pumped source. The measured single photon purity is $1 -g^2(0) = 76\%$, lower than that measured with free-spacing pumping. This decrease in purity is attributed to insufficient rejection of the back-reflected pump laser from the chip facet; a problem that can be remedied with an anti-reflection coating and/or a better FBG. 

In conclusion, we have demonstrated a hybrid integration method where single InAsP quantum dots can be placed anywhere within a silicon-based integrated photonic circuit and employed as a scalable single photon source. The method allows for pre-selected single quantum dots, epitaxially grown within InP nanowire waveguides, to be transfered and placed, at will, on SiN ridge waveguides. The tapered index profile of the nanowire, defined during growth, is used to evanescently couple to the waveguide system. The quantum dot photoluminescence can then be efficiently routed into the ridge waveguide and time-correlated measurements confirm the single photon nature of the coupled emission collected at the ridge waveguide facet. Measured coupling efficiencies were 74\%, whilst numerical studies predict values close to 100\% for an optimized nanowire taper geometry. Finally, we demonstrated that optical pumping and collection of the resulting quantum dot emission can be performed from the same ridge waveguide, allowing for attractive fiber-coupled solutions for plug-and-play applications using fixed, V-groove-type waveguide-fiber alignment techniques\cite{Shi_PTL2014}.

\textbf{Fabrication:} The dielectric films SiO$_2$ and SiN were deposited using plasma-enhanced chemical vapor deposition (PECVD) using SiH$_4$ chemistry. Gold alignment marks were defined on the first SiO$_2$ layer using e-beam lithography and lift-off. The SiN waveguides were defined after nanowire transfer using e-beam lithography and dry etching and positioned under the nanowires using the alignment marks. See Supplementary Information for additional details on device fabrication.

\textbf{Optical Measurements:} Optical studies were performed in a fiber-coupled continuous flow helium cryostat at 4K. Free-space laser excitation was directed normal to the nanowire through a 50X microscope objective (N.A. = 0.42). Excitation through the SiN waveguide was performed through a cleaved 780-HP fiber aligned in an endfire configuration to the waveguide facet at the sample edge. Excitation was above-band using both cw ($\lambda = 632.8$\,nm from a HeNe laser) and pulsed ($\lambda = 790$\,nm from a Ti-Sapphire laser, pulse width of 1.2\,ps, repetition rate of 80\,MHz). Collection was through the same fiber and dispersed using a 0.5\,m grating spectrometer and detected using a liquid-nitrogen cooled Si CCD for PL or filtered using a tunable FBG ($100\,\mu$eV resolution) and detected using two Si APDs (675\,ps timing jitter) for $g^2(\tau)$ measurements. 

\textbf{Modeling $g^2(\tau)$:} Our stochastic model of the excitonic emission process includes excitonic excitation due to a continuous chain of pump pulses with given period, followed by exciton emission with fitted lifetime and detection probability. Exciton re-excitation after a particular pulse due to the decaying band-edge carrier population was also included via a competition between carrier recombination time and dot occupation time.  A second-order correlation curve, $g^2(\tau)$, was generated by binning the resulting exciton emission times until a desired number of detection events was obtained. The calculated $g^2(\tau)$ curves were convolved with the measured detector response time of 675\,ps to compare with experiment.

This work was supported by the Canadian Space Agency.  

\bibliography{whiskers}
\end{document}